# Parallel Sparse Identification of Nonlinear Dynamics in Python: Task- and Data-Parallel SINDy with Numba, OpenMP, Threads and Multiprocessing for the Lorenz and Rössler Systems

Pablo Guillen-Rondon
*University of Houston–Downtown, Houston, TX, USA*
guillenrondonp@uhd.edu

**Abstract.** The sparse identification of nonlinear dynamics (SINDy) recovers interpretable governing equations from time-series data by sparse regression over a library of candidate functions. As data sets, candidate libraries and the number of repeated fits grow, the cost of SINDy becomes dominated by the evaluation of the library matrix and its least-squares problems. We present an open-source Python implementation of SINDy in which a single set of loop-based kernels is just-in-time compiled with Numba and executed on a shared-memory multicore node by four interchangeable back ends: Numba `prange`, OpenMP directives through PyOMP, Python threads that release the global interpreter lock, and `multiprocessing` with shared memory. The normal equations are accumulated in one cache-blocked pass over the data, with an automatic fallback to a tall-skinny QR reduction for ill-conditioned problems. Two decompositions are studied on the Lorenz and Rössler systems: an *ensemble* (task-parallel) workload of 960 independent identifications and a *library* (data-parallel) workload of one trajectory with $2\times10^7$ samples and a fifth-order library of 56 candidate terms. All back ends recover the exact sparsity pattern in every identification, with coefficient errors identical to PySINDy. On a dual-socket Intel Xeon E5-2650 v4 node of the University of Houston–Downtown HPC cluster (24 cores, 48 hardware threads), the compiled serial code is 4.3–5.6× faster than PySINDy, the parallel back ends reach speed-ups of 14.8–20.3 on 24 cores (parallel efficiency 0.61–0.84), and the overall time to solution is reduced by up to 117× with respect to PySINDy. We analyse the differences between the back ends, the effect of the two sockets and of hyper-threading, and give practical recommendations.

**Keywords:** system identification; SINDy; sparse regression; parallel computing; OpenMP; Numba; multiprocessing; high-performance Python; Lorenz system; Rössler system.

## 1. Introduction

Discovering the governing equations of a dynamical system directly from measurements is a central problem in data-driven science [9]. The sparse identification of nonlinear dynamics (SINDy) algorithm of Brunton, Proctor and Kutz [1] addresses it by assuming that the right-hand side of the dynamics is a sparse combination of a few candidate functions, so that a sparse regression over a library of such functions yields models that are both accurate and interpretable. SINDy has since been extended to partial differential equations [5], latent coordinates [6], implicit and rational dynamics [7], weak formulations [8] and ensemble/bootstrap variants for noisy data [4], with a recent tutorial perspective for engineering applications [11] and robust autoencoder variants [12]. The open-source PySINDy package [2, 3, 10] made these methods widely accessible.

Practical SINDy studies rarely consist of a single fit. Model selection requires sweeps over the sparsity threshold and the library; robustness is assessed by bootstrapping [4]; and long or high-resolution experiments produce millions of samples while high-order libraries contain tens to hundreds of terms. In this regime the computation is dominated by the evaluation of the library matrix $\Theta(X)$ and by least-squares solves, and interpreter overhead and serial execution become the limiting factors.

Python offers several mature routes to multicore execution: Numba's just-in-time (JIT) compilation with `prange` loops [15]; OpenMP [19] directives through PyOMP [16, 17] or pure-Python implementations such as OMP4Py [18]; ordinary threads, which run in parallel when compiled code releases the global interpreter lock (GIL) [21]; and process-based parallelism with `multiprocessing`. Their relative merits for equation-discovery workloads have not been systematically compared on a modern multicore node.

This paper makes three contributions. (i) We formulate SINDy so that its expensive part is a single cache-blocked pass that accumulates the normal equations, safeguarded by a tall-skinny QR (TSQR) fallback [27] for ill-conditioned data, and implement it as one set of compiled kernels shared by all back ends. (ii) We describe task-parallel (ensemble) and data-parallel (library) decompositions and their implementation with `prange`, PyOMP, GIL-free threads and `multiprocessing`. (iii) We report a strong-scaling study from 1 to 48 hardware threads on a compute node of the University of Houston–Downtown (UHD) HPC cluster for the Lorenz [13] and Rössler [14] systems, validated against PySINDy. All code and benchmark scripts are released as open source.

## 2. Sparse identification of nonlinear dynamics

We consider autonomous systems $\dot{x}(t) = f(x(t))$ with state $x \in \mathbb{R}^n$. Sampling the state at times $t_1, \ldots, t_N$ gives the data matrix $X \in \mathbb{R}^{N\times n}$ and, by numerical differentiation, $\dot{X} \in \mathbb{R}^{N\times n}$. SINDy postulates

$$\dot{X} \approx \Theta(X)\,\Xi, \tag{1}$$

where $\Theta(X) \in \mathbb{R}^{N\times m}$ collects m candidate functions evaluated on the data and $\Xi = [\xi_1, \ldots, \xi_n] \in \mathbb{R}^{m\times n}$ is a sparse coefficient matrix whose column $\xi_k$ defines the k-th equation $\dot{x}_k = \Theta(x)\xi_k$ [1]. Θ contains all monomials of the state variables up to degree d, ordered as 1, x, y, z, $x^2$, xy, xz, $y^2$, yz, $z^2$, $x^3$, … (as in PySINDy), so that m = C(n+d, d): m = 20 for d = 3 and m = 56 for d = 5 with n = 3.

### 2.1 Numerical differentiation

Derivatives are obtained with second-order central differences, $\dot{x}_i \approx (x_{i+1} - x_{i-1})/(2\Delta t)$, and second-order one-sided formulas at the two end points, matching `pysindy.FiniteDifference`. The truncation error is $O(\Delta t^2)$ and, for the noise-free data used here, is the dominant source of the coefficient errors reported in Section 5.

### 2.2 Sequentially thresholded least squares

The sparse coefficients are computed with sequentially thresholded least squares (STLSQ) [1], whose convergence was analysed in [26]. Starting from the full support, each iteration solves a ridge-regularised least-squares problem on the active columns and zeroes every coefficient whose magnitude falls below a threshold λ:

$$\xi_k \leftarrow \operatorname{argmin}_\xi \, ||\Theta_S\xi - \dot{x}_k||_2^2 + \alpha||\xi||_2^2, \quad S \leftarrow \{\, j : |\xi_{k,j}| \geq \lambda \,\}, \tag{2}$$

until the support S no longer changes. We use λ = 0.1 for Lorenz, λ = 0.05 for Rössler and $\alpha = 10^{-10}$. Columns are scaled to unit norm for the solve, while the ridge term and the threshold are applied in the original (unscaled) units, as in PySINDy.

### 2.3 Normal-equation formulation

Every least-squares problem in STLSQ involves sub-matrices of Θ only through the Gram matrix and the projected derivatives

$$G = \Theta^\top\Theta = \Sigma_{i=1}^N \theta_i\theta_i^\top, \quad R = \Theta^\top\dot{X} = \Sigma_{i=1}^N \theta_i\dot{x}_i^\top, \tag{3}$$

where $\theta_i^\top$ is the i-th row of Θ. The restricted problem for support S is $(G_{SS} + \alpha I)\xi_S = R_{S,k}$, an m×m system independent of N. A single pass over the data therefore costs $O(Nm^2)$ operations and $O(m^2)$ memory, after which all STLSQ iterations cost $O(m^3)$. The N×m matrix Θ is never stored: rows are generated in blocks of 4096 that fit in cache. Because G and R are sums over rows, the formulation is naturally parallel (Section 4). Since forming G squares the condition number of Θ [28], the code also checks the condition number of the small column-scaled Gram matrix and, above $10^{10}$, recomputes the fit from the R factor of a blocked tall-skinny QR factorisation of $[\Theta \mid \dot{X}]$ [27], which can be distributed in the same way; this happens only for nearly degenerate data.

## 3. Test systems and identified models

We use the two classical chaotic benchmarks of the SINDy literature [1]. The Lorenz system [13]

$$\dot{x} = \sigma(y - x), \quad \dot{y} = x(\rho - z) - y, \quad \dot{z} = xy - \beta z, \tag{4}$$

with σ = 10, ρ = 28, β = 8/3, and the Rössler system [14]

$$\dot{x} = -y - z, \quad \dot{y} = x + ay, \quad \dot{z} = b + z(x - c), \tag{5}$$

with a = b = 0.2, c = 5.7. Data are generated with a fourth-order Runge–Kutta (RK4) integrator with Δt = 0.002 (Lorenz) and Δt = 0.005 (Rössler) from random initial conditions in [−15,15]×[−20,20]×[5,40] and $[-8,8]^2\times[0,2]$, respectively. The data are noise-free, so that differences between back ends can only come from floating-point summation order.

Table 1 lists the coefficients identified in the library workload, from one trajectory of $N = 2\times10^7$ samples with the fifth-order library (56 candidate terms per equation, 168 in total). In both systems SINDy retains exactly the seven true terms and sets the other 161 coefficients to zero; the largest relative error is 0.16 %. Figure 1 compares simulations of the identified and true models from the same initial condition: the Rössler trajectories are indistinguishable, and the Lorenz trajectories coincide for about seven time units before separating, as expected from the positive Lyapunov exponent of the Lorenz attractor, while both remain on the same attractor.

**Table 1.** True and identified coefficients (library workload, $N = 2\times10^7$, degree-5 library). All other 161 coefficients are exactly zero in both systems. All parallel back ends give the same values to the digits shown.

| System | Eq. | Term | True | Identified | Rel. error |
|---|---|---|---|---|---|
| Lorenz | $\dot{x}$ | x | −10.0000 | −9.9991 | 0.009 % |
| | $\dot{x}$ | y | 10.0000 | 9.9991 | 0.009 % |
| | $\dot{y}$ | x | 28.0000 | 27.9920 | 0.029 % |
| | $\dot{y}$ | y | −1.0000 | −0.9984 | 0.156 % |
| | $\dot{y}$ | xz | −1.0000 | −0.9998 | 0.022 % |
| | $\dot{z}$ | z | −2.6667 | −2.6664 | 0.012 % |
| | $\dot{z}$ | xy | 1.0000 | 0.9999 | 0.012 % |
| Rössler | $\dot{x}$ | y | −1.0000 | −1.0000 | 0.000 % |
| | $\dot{x}$ | z | −1.0000 | −1.0000 | 0.004 % |
| | $\dot{y}$ | x | 1.0000 | 1.0000 | 0.001 % |
| | $\dot{y}$ | y | 0.2000 | 0.2000 | 0.001 % |
| | $\dot{z}$ | 1 | 0.2000 | 0.2000 | 0.013 % |
| | $\dot{z}$ | z | −5.7000 | −5.6993 | 0.013 % |
| | $\dot{z}$ | xz | 1.0000 | 0.9999 | 0.013 % |

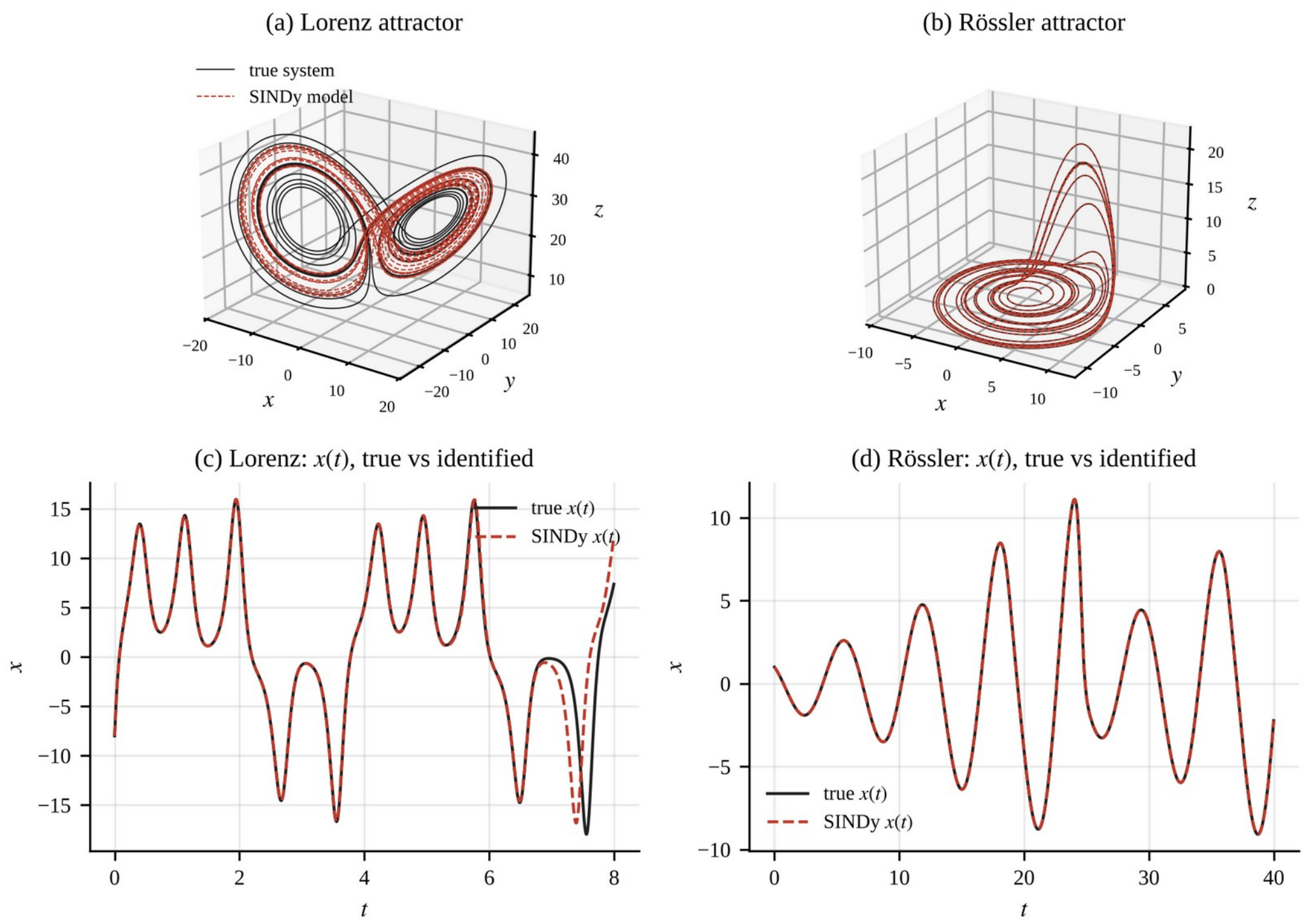

**Figure 1.** Identified versus true dynamics. (a, b) Attractors of the true system (solid black) and of the SINDy model (dashed red) integrated from the same initial condition. (c, d) Time series of x(t). For Lorenz the trajectories separate after $t \approx 7$ because of sensitive dependence on initial conditions, although the identified model reproduces the attractor; for Rössler they overlap for the whole window.

## 4. Implementation and parallelization strategies

### *4.1 Single-source compiled kernels*

All numerical steps (the RK4 integrator, finite differences, the library row $\theta_i$, accumulation of G and R, a pivoted Gaussian-elimination solver, the conditioning check, the QR fallback and STLSQ) are written once as plain loops in a factory `build_kernels(jit)` that receives the decorator to use. The same source is compiled with `numba.njit(nogil=True)` for the serial, `prange`, thread and process back ends and with PyOMP's `numba.openmp.njit` for OpenMP, so that every back end executes the same arithmetic and differences in run time are due to the parallelization strategy alone. Inner products are written as loops rather than BLAS calls, so that no hidden library threads distort the core count.

### *4.2 Two parallel decompositions*

Figure 2 summarises the pipeline and the two decompositions.

- **Ensemble (task parallelism).** M = 960 trajectories, each with n = 20 000 samples, are generated from different initial conditions and identified independently with a third-order library (m = 20). The work items are complete SINDy problems (integration, differentiation, library, STLSQ); workers do not communicate, and the result is M coefficient matrices. This pattern covers parameter sweeps, threshold searches, bootstrapping (ensemble-SINDy [4]) and multi-experiment studies.
- **Library (data parallelism).** A single trajectory with $N = 2\times10^7$ samples is identified with the fifth-order library (m = 56). Row blocks are distributed among p workers; worker q accumulates private partial sums $G_q$, $R_q$ over its rows, the partial sums are reduced, $G = \Sigma_q G_q$, and STLSQ is solved once. Only $O(pm^2)$ values are combined, independently of N.

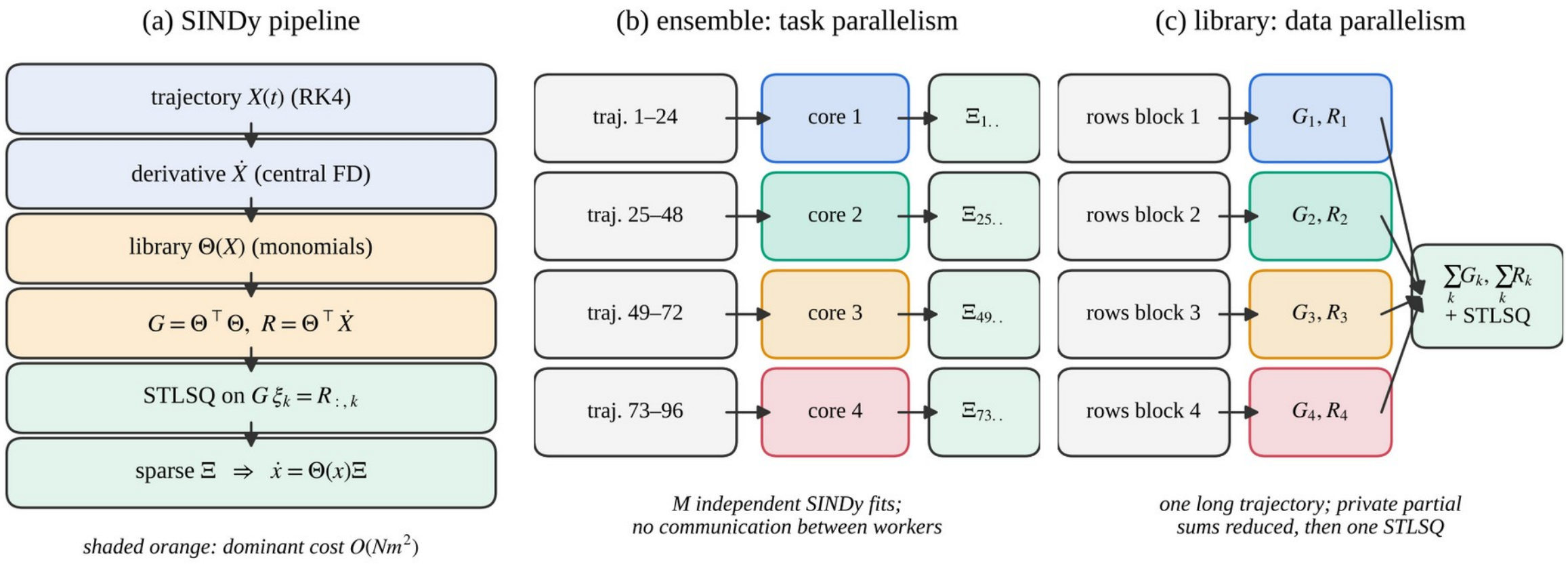

**Figure 2.** (a) SINDy pipeline in the normal-equation formulation; the orange steps dominate the cost. (b) Ensemble workload: whole identification problems are distributed among workers. (c) Library workload: row blocks of one long trajectory are distributed, private partial normal equations are reduced, and STLSQ is solved once.

### *4.3 Back ends*

**Numba `prange`.** A driver compiled with `@njit(parallel=True)` replaces `range` by `prange`, and Numba distributes the iterations over a pool of p native threads. In the library workload each iteration owns one slice of the private arrays `Gp[t]`, avoiding races without atomic operations:

```
@njit(parallel=True)
def ensemble_prange(system, p, x0s, dt, n, expo, thr):
    out = np.empty((x0s.shape[0], expo.shape[0], 3))
    for t in prange(x0s.shape[0]):          # one SINDy fit per iteration
        out[t] = sindy_one(system, p, x0s[t], dt, n, expo, thr)
    return out

@njit(parallel=True)
def library_prange_fit(X, dX, expo, thr, nparts):
    Gp = np.zeros((nparts, m, m)); Rp = np.zeros((nparts, m, 3))
    for t in prange(nparts):                # private partial sums
        gram_blocks(X, dX, expo, b0(t), b1(t), BLOCK, Gp[t], Rp[t])
    G = Gp.sum(axis=0); R = Rp.sum(axis=0)  # reduction
    return solve_checked(G, R, X, dX, expo, thr)   # STLSQ (+QR fallback)
```

**OpenMP (PyOMP).** PyOMP [16, 17] lowers `with openmp(...)` blocks to calls into the OpenMP runtime, giving Python code the semantics of C/Fortran `#pragma omp` directives, including explicit scheduling, thread identifiers and private data:

```
from numba.openmp import njit, openmp_context as openmp, omp_get_thread_num
@njit
def library_omp_fit(X, dX, expo, thr, nthreads):
    Gp = np.zeros((nthreads, m, m)); Rp = np.zeros((nthreads, m, 3))
    with openmp("parallel num_threads(nthreads)"):
        tid = omp_get_thread_num()
        with openmp("for schedule(static)"):
            for b in range(nblocks):
                library_gram_chunk(X, dX, expo, i0(b), i1(b), Gp[tid], Rp[tid])
    G = Gp.sum(axis=0); R = Rp.sum(axis=0)
    return solve_checked(G, R, X, dX, expo, thr)
```

The ensemble driver is a single `with openmp("parallel for schedule(static)")` loop over trajectories.

**Python threads.** Kernels compiled with `nogil=True` release the GIL for their whole duration, so a `concurrent.futures.ThreadPoolExecutor` with p threads achieves true parallelism while the driver remains ordinary Python code. The work is split into p contiguous chunks, so the GIL is acquired only once per chunk, and threads share all arrays.

**multiprocessing.** A `multiprocessing.Pool` of p worker processes, each with its own interpreter, executes the serial kernel on its chunk. In the ensemble only the initial conditions are sent and the M×m×3 coefficients returned. In the library workload X and $\dot{X}$ (0.96 GB) are copied once into `multiprocessing.shared_memory` blocks to which the workers attach at start-up, so each task transfers two block indices and returns an m×m and an m×3 matrix.

## 5. Performance evaluation

### *5.1 Experimental setup*

All measurements were taken on one compute node of the High-Performance Computing Center of the College of Sciences and Technology at UHD, a cluster of 32 compute nodes connected by Intel Omni-Path. The node has two Intel Xeon E5-2650 v4 processors (Broadwell, 2.2 GHz), each with 12 cores, for 24 physical cores and 48 hardware threads with Intel Hyper-Threading, and 256 GB of memory. The node was allocated exclusively through the Slurm scheduler. The software was Python 3.12.16 with NumPy 2.5.4, Numba 0.67.0, PyOMP and PySINDy 2.1.0 from conda-forge. Each configuration runs in a fresh interpreter in which the thread-pool size is fixed before Numba is imported and BLAS libraries are pinned to one thread. Each timing is the best of three repetitions after an untimed warm-up that triggers JIT compilation and starts the thread or process pools. The PySINDy reference uses the same data and settings (`STLSQ(threshold=λ, alpha=1e-10)`, `PolynomialLibrary(degree=d)`, `FiniteDifference()`) and is timed once. The worker counts p = 1, 2, 4, 8, 12, 16, 24 cover one socket (p ≤ 12) and both sockets (p = 24) with one thread per physical core, and p = 48 uses all hardware threads.

We report the speed-up and parallel efficiency with respect to the compiled serial code,

$$S(p) = T_{serial} / T(p), \quad E(p) = S(p) / p, \tag{6}$$

so that the gain due to compilation, reported separately, does not inflate the parallel speed-up.

### *5.2 Accuracy*

All back ends identify the same models. In the ensemble workloads all 960 Lorenz and all 960 Rössler identifications have the correct support, with maximum coefficient errors of $8.4\times10^{-3}$ and $2.1\times10^{-3}$ (medians $7.9\times10^{-3}$ and $7.2\times10^{-4}$); in the library workloads the maximum errors are $8.0\times10^{-3}$ and $7.2\times10^{-4}$. These values agree with those of PySINDy on the same data to at least four significant digits, confirming that the normal-equation formulation with the conditioning safeguard is as accurate as the reference implementation for these problems.

### *5.3 Run time and scaling*

Table 2 lists the timings, and Figure 3 shows the speed-up and the parallel efficiency.

**Table 2.** Wall time T [s] (best of three) and speed-up $S = T_{serial}/T$ on the UHD compute node (2 × Xeon E5-2650 v4, 24 cores, 48 hardware threads). Ensemble: M = 960, n = 20 000, degree 3; library: $N = 2\times10^7$, degree 5. For PySINDy the S column is $T_{serial}/T_{PySINDy}$. Rows with p = 24 (all physical cores) are shaded green, rows with p = 48 (hyper-threading) grey. Results for p = 2, 8 and 16 appear in Figure 3 and in the repository.

| Back end | p | L-ens T | S | L-lib T | S | R-ens T | S | R-lib T | S |
|---|---|---|---|---|---|---|---|---|---|
| PySINDy (serial) | 1 | 42.41 | 0.23 | 235.29 | 0.21 | 42.42 | 0.23 | 278.41 | 0.18 |
| Numba serial | 1 | 9.76 | 1.00 | 49.80 | 1.00 | 9.71 | 1.00 | 50.00 | 1.00 |
| Numba prange | 1 | 9.746 | 1.00 | 49.56 | 1.00 | 9.713 | 1.00 | 49.78 | 1.00 |
| | 4 | 2.546 | 3.83 | 14.17 | 3.51 | 2.510 | 3.87 | 14.42 | 3.47 |
| | 12 | 0.937 | 10.41 | 5.631 | 8.84 | 0.934 | 10.40 | 5.505 | 9.08 |
| | 24 | 0.489 | 19.96 | 2.822 | 17.65 | 0.493 | 19.68 | 2.850 | 17.54 |
| | 48 | 0.455 | 21.43 | 2.599 | 19.16 | 0.452 | 21.47 | 2.525 | 19.80 |

| Back end | p | L-ens T | S | L-lib T | S | R-ens T | S | R-lib T | S |
|---|---|---|---|---|---|---|---|---|---|
| PyOMP (OpenMP) | 1 | 10.14 | 0.96 | 50.44 | 0.99 | 10.13 | 0.96 | 50.39 | 0.99 |
| | 4 | 2.657 | 3.67 | 14.67 | 3.40 | 2.664 | 3.64 | 14.86 | 3.36 |
| | 12 | 0.975 | 10.01 | 5.849 | 8.51 | 0.976 | 9.95 | 5.694 | 8.78 |
| | 24 | 0.509 | 19.18 | 3.109 | 16.02 | 0.510 | 19.05 | 3.125 | 16.00 |
| | 48 | 0.470 | 20.77 | 2.759 | 18.05 | 0.471 | 20.62 | 2.745 | 18.22 |
| Threads (nogil) | 1 | 9.753 | 1.00 | 49.54 | 1.01 | 9.719 | 1.00 | 49.91 | 1.00 |
| | 4 | 2.543 | 3.84 | 12.80 | 3.89 | 2.516 | 3.86 | 12.85 | 3.89 |
| | 12 | 0.982 | 9.94 | 4.835 | 10.30 | 0.970 | 10.01 | 4.855 | 10.30 |
| | 24 | 0.576 | 16.93 | 2.469 | 20.18 | 0.526 | 18.45 | 2.469 | 20.25 |
| | 48 | 0.480 | 20.31 | 2.377 | 20.95 | 0.469 | 20.73 | 2.382 | 21.00 |
| multiprocessing | 1 | 9.774 | 1.00 | 49.57 | 1.00 | 9.783 | 0.99 | 49.53 | 1.01 |
| | 4 | 2.670 | 3.65 | 12.84 | 3.88 | 2.585 | 3.76 | 12.59 | 3.97 |
| | 12 | 0.980 | 9.96 | 4.887 | 10.19 | 1.031 | 9.42 | 4.932 | 10.14 |
| | 24 | 0.661 | 14.75 | 2.508 | 19.86 | 0.585 | 16.59 | 2.503 | 19.98 |
| | 48 | 0.482 | 20.25 | 2.384 | 20.89 | 0.479 | 20.27 | 2.408 | 20.77 |

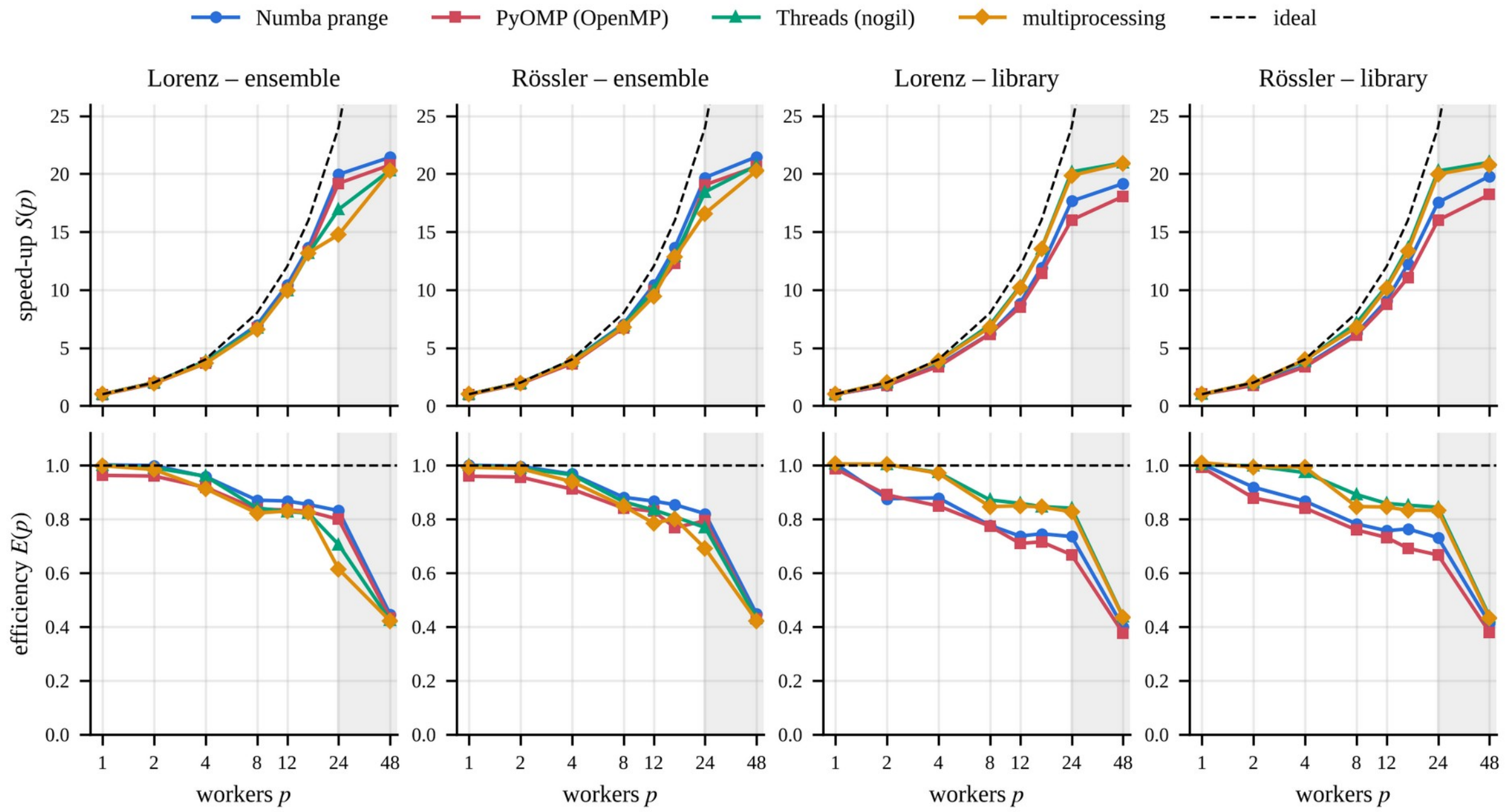

**Figure 3.** Strong scaling of the four back ends on the UHD node. Top: speed-up $S(p) = T_{serial}/T(p)$, with ideal linear scaling dashed. Bottom: parallel efficiency $E(p) = S(p)/p$. The worker axis is logarithmic; the shaded region ($p > 24$) uses hyper-threads.

**Compilation and formulation gain.** For identical results, the compiled serial code is 4.3× (Lorenz) and 4.4× (Rössler) faster than PySINDy in the ensemble workload and 4.7× and 5.6× faster in the library workload, where PySINDy takes 235 s and 278 s. In the library case the gain is partly algorithmic: PySINDy materialises the $2\times10^7\times56$ matrix (about 9 GB) and solves a tall least-squares problem in every STLSQ iteration, whereas the normal-equation formulation reads the data once and iterates on a 56×56 system.

**Scaling on physical cores.** All back ends scale well up to the 24 physical cores. At p = 12, one fully used socket, the efficiencies lie between 0.71 and 0.87; extending to the second socket at p = 24 costs little further efficiency for the best back end of each workload, which shows that the cache-blocked kernels are compute-bound rather than limited by memory bandwidth or NUMA effects. On 24 cores, Numba `prange` reaches S = 19.96 and 19.68 on the Lorenz and Rössler ensembles, and Python threads reach S = 20.18 and 20.25 on the library workloads. Combined with compilation, the best 24-core runs are 87× and 86× faster than PySINDy for the Lorenz and Rössler ensembles, and 95× and 113× faster for the library workloads, turning a 4.6-minute PySINDy fit into 2.5 s.

**Differences between back ends.** In the ensemble workload, the two compiled drivers (`prange` and PyOMP) perform best at p = 24 (S ≈ 19–20), whereas threads (S = 16.93 and 18.45) and `multiprocessing` (S = 14.75 and 16.59) fall behind. With a parallel run time of about 0.5 s, the fixed cost of dispatching work from Python to a thread or process pool and, for processes, of returning results through inter-process communication becomes a visible fraction of the total. In the library workload the order is reversed: threads and `multiprocessing` reach E(24) ≈ 0.83–0.84, while `prange` and PyOMP reach 0.74 and 0.67 for Lorenz. The gap is already present at p = 2 (S = 1.75 for `prange` against 2.00 for threads), so it is a per-thread overhead of the runtime-managed thread pools in this kernel rather than a scalability limit. A likely cause is the allocation of the per-block Θ buffer inside the parallel region; preallocating one buffer per thread is a simple remedy that we leave for a future version. The two compiled drivers, `prange` and PyOMP, behave almost identically throughout, which indicates that the choice between them can be made on grounds of programming style.

**Hyper-threading.** Doubling the workers from 24 to 48 hardware threads improves the run time only by a factor 1.04–1.37 (S(48) = 18.1–21.5), and the efficiency drops to 0.38–0.45. This is not explained by Amdahl's law [22]: fitting S(24) = 19.96 gives a serial fraction f ≈ 0.0088, which would predict S(48) ≈ 34. Instead, the two hardware threads of a core share its floating-point units, which the dense, compute-bound kernels already keep busy. The largest hyper-threading gains occur for the back ends with the most idle time at p = 24 (threads and processes in the ensemble), where the second hardware thread hides dispatch latency. For production runs on this node, 24 workers, one per physical core, are therefore the recommended setting.

## 6. Discussion and practical recommendations

- **Reformulate before parallelizing.** Accumulating $\Theta^\top\Theta$ and $\Theta^\top\dot{X}$ in a single cache-blocked pass reduces memory from O(Nm) to O($m^2$), makes STLSQ independent of N, exposes a simple reduction for data parallelism, and is responsible for a large part of the gain over PySINDy. The squared condition number is handled by checking the condition of the small Gram matrix and falling back to a TSQR reduction, which keeps the same parallel structure.
- **Choose the back end by task granularity.** Compiled drivers (`prange`, PyOMP) are best for many fine-grained tasks, because they dispatch work without returning to the interpreter. Python threads with `nogil` kernels and `multiprocessing` with shared memory are equally good or better for coarse tasks with little communication, and allow the parallel driver to remain ordinary Python code. `multiprocessing` additionally isolates memory and is the natural stepping stone to distributed execution.
- **Use physical cores.** On this node all back ends reach most of their attainable speed-up at 24 workers; hyper-threads add at most 37 % and usually much less.
- **OpenMP when explicit control matters.** PyOMP exposes scheduling, data-sharing clauses and thread identifiers with familiar OpenMP semantics and performed on par with `prange`, which eases porting from C/Fortran. The free-threaded CPython builds enabled by PEP 703 [21] may further simplify thread-based pipelines.

**Limitations and future work.** The experiments use noise-free data, polynomial libraries and a single shared-memory node. Noisy data require smoothing or weak formulations [8] and ensemble statistics [4], which increase the number of fits and therefore benefit directly from the ensemble strategy. Because both decompositions reduce to independent tasks or to a sum of small matrices, they extend naturally to distributed memory with MPI (a reduction of m×m Gram matrices or $\hat{R}$ factors across nodes), which we plan to evaluate across the 32 nodes of the UHD cluster, together with GPU back ends.

## 7. Conclusions

We presented a parallel implementation of SINDy in Python in which a single set of compiled kernels is executed by Numba `prange`, PyOMP, GIL-free threads and `multiprocessing`. For the Lorenz and Rössler systems all back ends recover the exact governing equations in every one of the 960 ensemble identifications and in the $2{\times}10^7$-sample library fits, with errors identical to PySINDy. On a 24-core compute node of the UHD HPC cluster, the normal-equation formulation and JIT compilation make the serial code 4.3–5.6× faster than PySINDy, and parallel execution adds speed-ups of up to 20.3 on the physical cores, for an overall reduction of the time to solution of up to 117×. Task-parallel ensembles and data-parallel library construction thus make SINDy studies with large data sets, rich libraries and many repeated fits practical on a single multicore node.

## Acknowledgments

The author gratefully acknowledges the High-Performance Computing Center of the College of Sciences and Technology at the University of Houston–Downtown, operated by UHD Information Technology, for providing the computational resources used in this work. The UHD HPC cluster was established with support from a grant of the U.S. Department of Defense.

## Code availability

The source code (kernels, back ends, benchmark driver, Slurm job script, raw results in JSON format and the scripts that generate every figure and table of this paper) is available at https://github.com/guillenrondonp/parallel-sindy under an open-source license. The benchmark is reproduced with `python benchmark.py --cores 1 2 4 8 12 16 24 48 --M 960 --N 20000000`.